\documentclass[showpacs,amsmath,amssymb,twocolumn,pra,superscriptaddress]{revtex4-2}

\usepackage{amsfonts,amsthm,mathrsfs,eufrak,xspace,framed}
\usepackage{xcolor}
\usepackage{algorithm}
\floatname{algorithm}{Protocol}
\usepackage{algorithmic}
\usepackage{subfigure}
\usepackage{amscd}
\usepackage{multirow}
\usepackage{graphicx}% Include figure files
\usepackage{dcolumn}% Align table columns on decimal points
\usepackage{bm}% bold math
\usepackage{bbm}% bold math
 \usepackage{enumitem}
 \usepackage{pbox}

\newcommand{\bra}[1]{\mbox{$\left\langle #1 \right|$}}
\newcommand{\ket}[1]{\mbox{$\left| #1 \right\rangle$}}

\newcommand{\be}{\begin{equation}}
\newcommand{\ee}{\end{equation}}
\newcommand{\bea}{\begin{eqnarray}}
\newcommand{\eea}{\end{eqnarray}}

\definecolor{mygreen}{rgb}{0,0.5,0}
\definecolor{myblue}{rgb}{0,0,0.75}
\definecolor{mymagenta}{cmyk}{0,1,0,0.12}

\usepackage{umoline}

\begin{document}
\title{Sensitivity Bounds of Multiparameter Metrology at Thermal Equilibrium}
\author{Zhu Cao}
\email{caozhu@tongji.edu.cn}
\affiliation{College of Electronics and Information Engineering, Tongji University, Shanghai 201804, China}
\affiliation{Shanghai Research Institute for Intelligent Autonomous Systems, Tongji University, Shanghai 201210, China}

\begin{abstract}
Quantum metrology aims to enhance measurement precision beyond the classical limit by leveraging quantum resources. Unlike multi-parameter dynamic quantum metrology, many questions regarding multiparameter quantum metrology at thermal equilibrium remain elusive. In particular, the ultimate precision limits achievable in this equilibrium setting are not yet well understood. In this work, we examine the fundamental limits of estimating multiple parameters with a quantum probe at thermal equilibrium. We first show that the Heisenberg limit with respect to the number of probes can be achieved, and our bound coincides with the known single-parameter bound when only one parameter is estimated. We then consider the low temperature limit, revealing a qualitatively different behavior compared to the finite temperature case. We give an example to illustrate the usage of our main results. Finally, we show the conditions under which the sensitivity bound can be attained and the optimal measurements to achieve it.
\end{abstract}

\maketitle

%\section{Introduction}

% Motivate quantum metrology: TO DO

\section{Introduction}

Quantum metrology \cite{giovannetti2004quantum,giovannetti2006quantum,degen2017quantum,giovannetti2011advances,demkowicz2014using,demkowicz2012elusive,huelga1997improvement,chin2012quantum,hall2012does,berry2015quantum,alipour2014quantum,beau2017nonlinear} is a rapidly developing field at the intersection of quantum physics and precision measurement science, dedicated to exploiting uniquely quantum phenomena to enhance the sensitivity of measurements beyond the limits imposed by classical physics. Applications of quantum metrology span a wide range of domains: improving the accuracy of gravitational wave detectors \cite{schnabel2010quantum,tse2019quantum}, advancing magnetic resonance imaging at the nanoscale \cite{brida2010experimental,le2013optical}, refining frequency standards in next-generation atomic clocks \cite{buvzek1999optimal,pedrozo2020entanglement}, and even enabling possible extensions of the Standard Model of particle physics \cite{bass2024quantum}. Furthermore, quantum metrology provides a bridge between fundamental and applied science, linking abstract concepts such as entanglement entropy \cite{eisert2010colloquium} and Fisher information \cite{liu2020quantum} to tangible experimental outcomes. In doing so, it also drives innovation in quantum technologies more broadly, including quantum communication \cite{gisin2007quantum} and computation \cite{divincenzo1995quantum}. As experimental control over quantum systems continues to advance, quantum metrology stands as one of the most promising avenues for translating the counterintuitive principles of quantum mechanics into transformative real-world capabilities \cite{aslam2023quantum}.

Currently, the most studied quantum metrology is dynamic metrology \cite{giovannetti2004quantum}. In single-parameter dynamic metrology, an initial $N$-partitequantum state $\rho_0$ undergoes a quantum channel with parameter $\theta$ for a period of time $t$, after which a measurement is performed. The parameter is then inferred from the measured statistics.
For single-parameter dynamic metrology, its fundamental sensitivity limit has been established in Ref.~\cite{giovannetti2006quantum}  as  
$F \le O(N^2 t^2)$,
where $F$ is the quantum Fisher information (QFI).
This quadratic dependence on $t$ characterizes the ideal noiseless case; under Lindblad-type noise, the QFI typically grows at most linearly in 
$t$ for short times \cite{zhou2018achieving}.
For multi-parameter dynamic metrology, its fundamental limit is also understood \cite{gessner2018sensitivity}.
Specifically, for an arbitrary real vector $\mathbf{n} $ that satisfies $||\mathbf{n}  ||_2=1$ (i.e., its 2-norm is 1), we have 
$\mathbf{n}^T \mathbf{F} \mathbf{n} \le O(N^2 t^2)$,
where $\mathbf{F}$ is the quantum Fisher information matrix.

Another important type of quantum metrology is equilibrium metrology. In this setting, an $N$-partite Gibbs
state is prepared with respect to a system Hamiltonian $H$ at inverse temperature $\beta$, thus encoding the Hamiltonian parameters. 
For gapped Hamiltonians, such a Gibbs state can be obtained via passive thermalization, whereas for gapless Hamiltonians, its preparation generally requires active control (see for example the method in Ref.~\cite{chen2025efficient}).
The Gibbs state is subsequently measured, the outcome of which is then utilized to estimate the parameters of interest~\footnote{There are also other types of quantum metrology in addition to the two mentioned. For example,  non-equilibrium thermal metrology uses a thermal state as the probe and encodes the parameter via unitary time evolution \cite{zhang2025universal}.}.
We note a crucial difference between dynamic metrology and equilibrium metrology.
In dynamic metrology, the parameter is imprinted through unitary evolution, so the distinguishability of states grows quadratically with the interrogation time $t$. In equilibrium metrology, by contrast, the probe is assumed to have fully thermalized to the Gibbs state $\rho = e^{-\beta H}/\mathrm{Tr}(e^{-\beta H})$. The parameter dependence is encoded in the static structure of the Hamiltonian within this equilibrium state rather than accumulated dynamically over a controllable evolution time. Consequently, $t$ does not constitute a metrological resource; instead, the relevant energy scale is set by the inverse temperature $\beta$, which plays a role analogous to $t$ in dynamic scenarios.

Recently, the fundamental precision limit of single-parameter equilibrium metrology has been established as $F \le O( \beta^2 N^2  )$ \cite{abiuso2025fundamental}.
In contrast, the multi-parameter case has received comparatively little attention \cite{mihailescu2024multiparameter,mihailescu2025metrological}. 
In Ref.~\cite{mihailescu2024multiparameter}, multi-parameter estimation at finite temperature is performed through the two-impurity Kondo model. Ref.~\cite{mihailescu2025metrological} establishes a link between metrological symmetries and the number of effective parameters that can be estimated in 
quantum equilibrium metrology. Despite these progresses, the fundamental precision limit for multi-parameter equilibrium metrology is poorly understood. In particular, 
its scaling with the number of probes $N$ and the inverse temperature $\beta$ remains elusive.

In this work, we fill this gap by establishing 
the fundamental precision limits of multi-parameter equilibrium metrology. 
Let the parameter vector be $\Theta=(\theta_1, \cdots, \theta_M)$. 
The $(j,k)$-th entry of the parameter covariance matrix $\Sigma$ is defined by
$ \Sigma_{jk}  =  \textrm{Cov}(\theta_j, \theta_k)$.
A modified quantum Cram\'er-Rao bound (QCRB) dictates 
$\mathbf{n}^T \Sigma \mathbf{n} \ge 1/ \mathbf{n}^T \mathbf{F} \mathbf{n}$ \cite{gessner2018sensitivity},
for any real $M$-dimensional vector $\mathbf{n} $ that satisfies $ ||\mathbf{n}||_2=1$ and that the scalar $\mathbf{n}^T \mathbf{F} \mathbf{n}$ is positive.
Note that this bound remains valid even when $\mathbf{F}$ is degenerate, a regime in which the standard quantum Cramér–Rao bound ceases to apply. 
We show that
\begin{equation}
\mathbf{n}^T \mathbf{F} \mathbf{n} \le O( \beta^2 N^2  ),
\end{equation}
for finite inverse temperature $\beta$, and
\begin{equation}
\mathbf{n}^T \mathbf{F} \mathbf{n} \le O( N^2/\Delta^2 ),
\end{equation}
in the zero-temperature limit $\beta \to \infty$, where $\Delta$ is the energy gap of the Hamiltonian $H$.
To showcase the power of these bounds, we provide a concrete example that 
compares the exact quantum Fisher information with the derived sensitivity bound. 
We also examine the attainability of the bounds and give concrete
examples in which they are saturated. Finally, we identify the optimal measurements
that can achieve the maximal Fisher information for a given thermal state.
A summary of the results of our work and related work is shown in Table \ref{tab:compare}.

\begin{table}[htbp]
    \centering
    \caption{
   A collection of sensitivity bounds for both dynamic metrology and equilibrium metrology, for both the 
   case of single-parameter estimation and that of multiple-parameter estimation. Our work fills the last missing
   piece of the table.
    }
    \renewcommand\arraystretch{2}
    \tabcolsep=2pt
    \vspace{5pt}
        \begin{tabular}{  ccc}
        \hline
        \hline        
        & Dynamic metrology &  Equilibrium metrology   \\
        \hline    
        \vspace{0.2cm}    
        
 \pbox[l]{2cm}{  Single parameter }&   
\pbox[l]{2.5cm}{ $F \le O(N^2 t^2)$   \cite{giovannetti2006quantum} } &  \pbox[l]{3cm}{  $F \le O( \beta^2 N^2  )$ \cite{abiuso2025fundamental} }\\

   \vspace{0.1cm}
    \pbox[l]{2cm}{ Multiple parameters }&  \pbox[l]{2.5cm}{ $\mathbf{n}^T \mathbf{F} \mathbf{n} \le O(N^2 t^2 )$  \cite{gessner2018sensitivity} }&
    \pbox[l]{3cm}{ $\mathbf{n}^T \mathbf{F} \mathbf{n} \le O( \beta^2 N^2  )$ (This work) } \\          
        \hline 
        \hline       
        \end{tabular}
    \label{tab:compare}
\end{table}

\section{Multiparameter Estimation via Equilibrium States}

In this section, we present the quantum Fisher information matrix for multiparameter estimation 
using equilibrium states. This provides the basis for later analysis of the
sensitivity bounds.

For multi-parameter estimation \cite{paris2009quantum}, 
the $(\mu,\nu)$-th entry of the quantum Fisher information matrix $\mathbf{F}=(F_{\mu,\nu})_{\mu,\nu \in [M]}$
has the form 
\begin{equation}
\label{eq:QFI}
F_{\mu,\nu} = {\rm Tr} \left[  \dot{\rho_\mu} J_{B,\rho}^{-1} [  \dot{\rho_\nu}  ]   \right],
\end{equation}
where $\dot{\rho_\mu}  \equiv \partial \rho / \partial \mu$ and likewise for $ \dot{\rho_\nu}$.
The superoperator $J_{B,\rho}^{-1}$ is defined as
$J_{B,\rho}^{-1}[A] = 2 \int_0^\infty ds e^{-\rho s} A e^{-\rho s}$
and is the inverse operator of $J_{B,\rho}$, which acts as
$J_{B,\rho}[A] = (\rho A + A \rho)/2$ \cite{scandi2023quantum}. For completeness, we provide a proof of this inverse operator 
relation in Appendix \ref{app:A}.

Substituting the Gibbs state 
$\rho = e^{- \beta H} / {\rm Tr} [ e^{-\beta H} ]$,
we obtain
\begin{equation}
\label{eq:thermal}
\beta^{-2} F_{\mu,\nu} = {\rm Tr} [ \dot{H_\mu} \mathcal{J}_\rho [ \dot{H_\nu}]  ] -  {\rm Tr}[  \rho \dot{H_\mu}  ] \rm{Tr} [ \rho \dot{H_\nu}  ].
\end{equation}
Here, $ \dot{H_\mu} \equiv \partial H / \partial \mu$, and likewise for $ \dot{H_\nu}$.
The operator $\mathcal{J}_\rho$  is defined as
$ \mathcal{J}_\rho = J_{L,\rho} \circ J_{B,\rho}^{-1} \circ J_{L,\rho}$,
where $J_{L,\rho} [A] = \int_0^1 ds \rho^s A \rho^{1-s}$.
The detailed derivation of Eq.~\eqref{eq:thermal} can be found in Appendix \ref{app:B}.

Equation~\eqref{eq:thermal} is the first main result of this work, which makes explicit that, at thermal equilibrium, the quantum Fisher information matrix is entirely determined by fluctuations of the Hamiltonian derivatives with respect to the parameters. The second term subtracts the product of expectation values, so that $F_{\mu,\nu}$ is governed by the connected correlations of $\dot H_\mu$ and $\dot H_\nu$. In this sense, the QFI matrix can be viewed as a generalized covariance matrix evaluated with respect to the Kubo–Mori–Bogoliubov inner product defined by $\mathcal{J}_\rho$. This formulation highlights two key features of equilibrium metrology: (i) sensitivity is controlled by thermal response functions rather than dynamical phase accumulation, and (ii) the geometric structure of the parameter space is encoded in equilibrium susceptibilities. These observations will be central to deriving the sensitivity bounds in the following sections.

\section{Bounds for Multiparameter Fisher Information Matrix at Thermal Equilibrium}

In this section, we present upper bounds of quantum Fisher information (QFI) matrix for multi-parameter 
estimation at thermal equilibrium.

In the rest of the paper, we consider a Hamiltonian that encodes the parameters locally,
\begin{equation}
H  =  \sum_{m=1}^M  (  \sum_{k=1}^{N_m}  h_m^{(k)} )  \theta_m,
\end{equation}
where $h_m^{(k)}$ acts on different qubits in the system and $\theta_m$ are the parameters. 
For simplicity, we assume that all operators $h_m^{(k)}$ share the same spectrum, with eigenvalues $\pm 1/2$, and
are identical up to their action on different subsystems. It is important note that 
 this assumption can be relaxed to 
that each operator $h_m^{(k)}$ are spectrally bounded by a constant without affecting the main results.

To give the bound for the QFI matrix, we first introduce the covariance matrix $\Gamma(\rho, H)$ associated with the state $\rho$ and the Hamiltonian $H$.
Its $(\mu,\nu)$-th entry 
is defined as $\Gamma(\rho, H)_{\mu,\nu} ={\rm Tr}[  \rho ( \dot{H_\mu} \dot{H_\nu} + \dot{H_\nu} \dot{H_\mu} )/2 ] -  {\rm Tr}[  \rho \dot{H_\mu}  ] \rm{Tr} [ \rho \dot{H_\nu}  ]$.
In Appendix \ref{app:C}, we show that, for any real column vector $\mathbf{n}=(n_1,\cdots, n_M)$ that satisfies $|| \mathbf{n} ||_2 =1$, the following inequality holds
\begin{equation}
\label{eq:thermal_bound}
\mathbf{n}^T \mathbf{F}(\rho, H) \mathbf{n}  \le  \beta^2 \mathbf{n}^T   \Gamma(\rho, H) \mathbf{n},
\end{equation}
which provides an upper bound on the QFI matrix.

To give a more accessible bound, we next examine the quadratic form $\mathbf{n}^T  \Gamma(\rho, H) \mathbf{n}$.
By maximizing over all quantum states $\rho$, we find 
\begin{equation}
\label{eq:GammaHL}
 \beta^2 \mathbf{n}^T   \Gamma(\rho, H) \mathbf{n} \le \beta^2 \mathbf{n}^T   \Gamma_{HL} \mathbf{n},
\end{equation}
where  $\Gamma_{HL} = \bold{v} \cdot \bold{v}^T$/4, with $\bold{v} =( \epsilon_1 N_1, \cdots, \epsilon_M N_M)$ and $\epsilon_k = \textrm{sgn}(n_k)$.
The upper bound is saturated by the quantum state $\rho=\ket{\psi}\bra{\psi}$, where
\begin{equation}
\label{eq:GammaSat}
\begin{aligned}
\ket{\psi} = &  \frac{1}{2} ( \ket{N_1, \epsilon_1} \otimes \ket{N_2, \epsilon_2}  \otimes \cdots \otimes \ket{N_M, \epsilon_M}  \\
  & +  \ket{N_1, -\epsilon_1} \otimes \ket{N_2, -\epsilon_2}  \otimes \cdots \otimes \ket{N_M,  -\epsilon_M}).
\end{aligned}
\end{equation}
Here, $\ket{N_m, 1}$ ($\ket{N_m, -1}$) denotes a tensor product of $N_m$ eigenstates of $h_m$, each with eigenvalue value $1/2$ ($-1/2$).
This is a GHZ state of $M$ large collective spins, aligned along a direction $\epsilon$ in parameter space, simultaneously extremal for all generators. When $M=1$, this state reduces to the normal cat state.
The derivation of Eq.~\eqref{eq:GammaHL} and Eq.~\eqref{eq:GammaSat} can be found in Appendix \ref{app:GammaOpt}.

For the single-parameter case ($M=1$), the bound reduces to
$F \le \beta^2 N^2 /4$,
which is consistent with the known single-parameter QFI bound presented in Ref.~\cite{abiuso2025fundamental}, taking $|| h || = \max_{m} || h_m || = 1$.
For $M>1$ and $N_1=\cdots=N_M=N$, the bound implies 
\begin{equation}
\label{eq:finitebound}
\mathbf{n}^T \Sigma \mathbf{n} \ge O(1/\beta^2 N^2),
\end{equation}
which is the second main result of this work.  In Appendix \ref{app:CV}, we extend this result to the continuous-variable case.

%\section{Low Temperature Limit}
\section{Low temperature limit}

It can be observed that the bound Eq.~\eqref{eq:finitebound} becomes trivial in the zero-temperature limit, i.e., when the inverse temperature satisfies $\beta = \infty$. Therefore we present in this section an alternative non-trivial sensitivity bound 
applicable at zero temperature.

Let $\ket{i}$ be an eigenstate of the Hamiltonian $H$, satisfying
$ H \ket{i} = E_i \ket{i}$,
where $E_i$ is the corresponding eigenvalue.
The Gibbs state associated with $H$ is given by
$\rho = \sum p_i \ket{i}\bra{i}$,
with $p_i = e^{-\beta E_i } / (\sum_j  e^{-\beta E_j} )$.
Introducing the notation $a:=p_i$ and $\Pi_a:=\ket{i}\bra{i}$, the density operator can be written compactly as
$\rho  = \sum_a  a \Pi_a$.

Using this notion, one finds
\begin{equation}
\label{eq:middle}
\begin{aligned}
 {\rm Tr} [B \mathcal{J}_\rho [ A ]  ]  =  & \sum_a  a {\rm Tr}[B \Pi_a A \Pi_a]   \\
 & +  \sum_{a \not= b} \frac{ 2(a-b)^2 }{ ( \ln a - \ln b)^2 (a+b) }  {\rm Tr}[B \Pi_b A \Pi_a].
 \end{aligned}
\end{equation}
The derivation of this identity can be found in Appendix \ref{app:D}.
Substituting Eq.~\eqref{eq:middle} into the expression of $\beta^{-2} F_{\mu,\nu} $, 
we obtain 
\begin{equation}
\begin{aligned}
\beta^{-2} F_{\mu,\nu} =  & {\rm Tr} [ \sum_a a ( \Pi_a  \dot{H_\mu}  \Pi_a) (\Pi_a  \dot{H_\nu}  \Pi_a) ]  \\
 & - {\rm Tr} [ \sum_a a \Pi_a \dot{H_\mu} \Pi_a ] {\rm Tr} [ \sum_a a \Pi_a \dot{H_\nu} \Pi_a ] \\
 & + \sum_{a \not= b} \frac{ 2(a-b)^2 }{ ( \ln a - \ln b)^2 (a+b) } {\rm Tr}[\dot{H_\mu} \Pi_b \dot{H_\nu} \Pi_a] .
 \end{aligned}
\end{equation}

Let $H^e = \sum_{i=1}^M n_i \dot{H_i} $. 
Then, the quadratic form of the QFI matrix can be written as
\begin{equation}
\beta^{-2} \mathbf{n}^T \mathbf{F} \mathbf{n} = Var(H_e) + \sum_{i \not = j} \frac{2(p_i - p_j)^2 | H_{ij}^e |^2  }{ (\ln p_i -\ln p_j)^2 (p_i + p_j) } ,
\end{equation}
where $H_{ij}^e$ is the $(i,j)$-th entry of $H^e$.
Writing $\beta^2 Var(H_e)$ as $F^{diag}$, we have
\begin{equation}
\mathbf{n}^T \mathbf{F} \mathbf{n} = F^{diag} + \sum_{i<j} \frac{4(p_i - p_j)^2 | H_{ij}^e |^2  }{ (E_i - E_j)^2 (p_i + p_j) }.
\end{equation}
Here, $p_i = exp(-\beta E_i)$, with $E_i$ the eigenvalue of $H^e$. Let $\Delta = E_1 - E_0$ denote the energy gap.
In the limit of large $\beta$, we find
\begin{equation}
\begin{aligned}
F^{diag}  = O( e^{-\beta \Delta}), \quad \quad\quad \quad & \\
 \sum_{0<i<j} \frac{4(p_i - p_j)^2 | H_{ij}^e |^2  }{ (E_i - E_j)^2 (p_i + p_j) }  =  O( e^{-\beta \Delta}). &
\end{aligned}
\end{equation}
Consequently,
 \begin{equation}
\mathbf{n}^T \mathbf{F} \mathbf{n}= \sum_{i>0} \frac{4 |H_{0i}^e |^2 }{ (E_i - E_0)^2}  + O( e^{-\beta \Delta} ).
 \end{equation}
By suitable manipulation of the inequalities, we have
\begin{equation}
\mathbf{n}^T \mathbf{F} \mathbf{n} \le \frac{|| H^e ||^2 }{ \Delta^2} + O( e^{-\beta \Delta} ).
\end{equation}

To make the $N$-scaling apparent, we now analyze the norm of $\|H^e\|$. 
Each operator $\dot{H}_i$ is a sum of $O(N)$ local terms,
\begin{equation}
\dot{H}_i = \sum_{k=1}^{N} h_{i}^{(k)},
\end{equation}
where each $h_{i}^{(k)}$ acts nontrivially only on a constant number of sites and satisfies 
$\|h_i^{(k)}\| = O(1)$. 
By subadditivity of the operator norm,
\begin{equation}
\|\dot{H}_i\| 
\le 
\sum_{k=1}^{N} \|h_i^{(k)}\| 
= O(N).
\end{equation}
Since $H^e = \sum_i n_i \dot{H}_i$ contains only a finite number $M$ of such operators 
(independent of $N$), we further obtain
\begin{equation}
\|H^e\| 
\le 
\sum_{i=1}^{M} |n_i| \, \|\dot{H}_i\| 
= O(N).
\end{equation}

Substituting the norm of $H^e$, we have
\begin{equation}
\mathbf{n}^T \mathbf{F} \mathbf{n} \le O( \frac{N^2}{ \Delta^2}).
\end{equation}
If the many-body energy gap $\Delta$ remains finite as the system size $N$ increases (i.e., away from criticality), the bound implies Heisenberg scaling with respect to $N$. By contrast, near a quantum phase transition the gap typically closes as $\Delta \sim N^{-z}$ in a finite system at criticality. In this case,
\begin{equation}
\mathbf{n}^T \mathbf{F} \mathbf{n} \lesssim O\!\left(N^{2+2z}\right),
\end{equation}
leading to apparent super-Heisenberg scaling with respect to $N$.
This enhancement originates from the diverging correlation length and critical fluctuations, which amplify the susceptibility of the state to small parameter variations. In particular, the quantum Fisher information is closely related to the fidelity susceptibility, a quantity known to exhibit universal scaling behavior near critical points. Such gap-enhanced precision is commonly referred to as \emph{critical quantum metrology}~\cite{zanardi2006ground,you2007fidelity,gu2010fidelity,rams2011quantum,frerot2018quantum,adani2024critical}.

We emphasize that this super-Heisenberg scaling does not contradict fundamental metrological bounds. Rather, it reflects the fact that the closing gap itself depends on $N$, so that the effective generator norm grows superlinearly with system size. When all relevant physical resources (including interaction strength, preparation time, and proximity to criticality) are properly accounted for, the overall scaling remains consistent with generalized Heisenberg-type limits.

\section{Example}

We illustrate the above general results with an example.
Consider the Ising Hamiltonian 
$H = -J \sum_{j=1}^{N-1} \sigma^j \cdot    \sigma^{j+1}    + \sum_{j=1}^N B^j \sigma_x^j$,
where $\sigma^j = (\sigma_x^j, \sigma_y^j,\sigma_z^j )$ are Pauli matrices.
The parameters to be estimated are the local fields $B=(B^1, \cdots, B^N)$.
Note that in this model, the condition $[\dot{H_i}, \dot{H_j}] =0$ is satisfied. 
Moreover, we have $|| h || = 1-(-1)=2$.
Therefore, the upper bound on the QFI reads
\begin{equation}
\mathbf{n}^T \mathbf{F}(\beta, H) \mathbf{n} \le \frac{\beta^2 N^2 || h ||^2  }{4} = \beta^2 N^2.
\end{equation}
Hence, the covariance is lower bounded by
$\mathbf{n}^T \textsf{Cov}(B)  \mathbf{n} \ge 1/\mathbf{n}^T \mathbf{F}(\beta, H) \mathbf{n}  \ge 1/N^2 \beta^2$.

We now compute the exact value of QFI for the Ising model.
The Hamiltonian can be written as 
\begin{equation}
H = - J \sum_{i=1}^{2N} \sigma_z^i \sigma_z^{i+1} +  B_1 \sum_{i=1}^N \sigma_z^{2i+1} + B_2 \sum_{i=1}^N  \sigma_z^{2i+2},
\end{equation} 
where the site indices are understood cyclically modulo $2N$. Let $\mu=B_1$, $\nu=B_2$.
First, we note that
\begin{equation}
\label{eq:Fexp}
F_{\mu \nu} = \frac{\partial^2}{\partial \mu \partial \nu} \ln \mathcal{Z},
 \end{equation}
 where $\mathcal{Z}$ denotes the partition function. A derivation of Eq.~\eqref{eq:Fexp} is shown in Appendix \ref{app:E}.

Hence, we can use the partition function $\mathcal{Z}$ to express the quantum Fisher information. The partition function $\mathcal{Z}$ admits the expansion
\begin{equation}
\begin{aligned}
&\mathcal{Z}
= \operatorname{Tr}\!\left[e^{-\beta H}\right] \\
=\,& e^{\beta J z_{2N} z_1} e^{-\beta B_1 z_1 }   e^{\beta J z_1 z_2 }  
   e^{-\beta B_2 z_2 } \cdots e^{-\beta B_2 z_{2N}}.
\end{aligned}
\end{equation}
To simplify this expression, let 
\begin{equation}
\begin{aligned}
A &=
\begin{pmatrix}
e^{-\beta B_1/2} & 0 \\
0 & e^{-\beta B_1/2 }
\end{pmatrix},
\quad
C =
\begin{pmatrix}
e^{\beta J} & e^{-\beta J} \\
e^{-\beta J} & e^{\beta J}
\end{pmatrix}, \\
D &=
\begin{pmatrix}
e^{-\beta B_2/2}  & 0 \\
0 & e^{-\beta B_2/2}\\
\end{pmatrix}. \\
\end{aligned}
\end{equation}     
Using this notation, the partition function can be rewritten as
\begin{equation}
\label{eq:Zexp}
\begin{aligned}
 \mathcal{Z} =& \sum\limits_{z_1, \cdots, z_{2N}=1,2}(DCA)_{z_{2N}z_1}(ACD)_{z_1z_2} \cdots (ACD)_{z_{2N-1}z_{2N}} \\
= &Tr[ (ACD)^{2N} ] = \lambda_{+}^{2N} + \lambda_{-}^{2N},
 \end{aligned}
 \end{equation}
where
$\lambda_\pm = e^{\beta J } (\cosh (\theta) \pm \sqrt{\sinh^2(\theta)+ e^{-4\beta J} } )$
 and $\theta = \beta (B_1 + B_2)/2$.
Substituting Eq.~\eqref{eq:Zexp} into Eq.~\eqref{eq:Fexp}, we obtain
\begin{equation}
F_{\mu \nu} = \frac{\partial^2}{\partial \mu \partial \nu} \ln (\lambda_{+}^{2N} + \lambda_{-}^{2N}). 
\end{equation}

\begin{figure}[htbp]
\centering
\includegraphics[width=8.5cm]{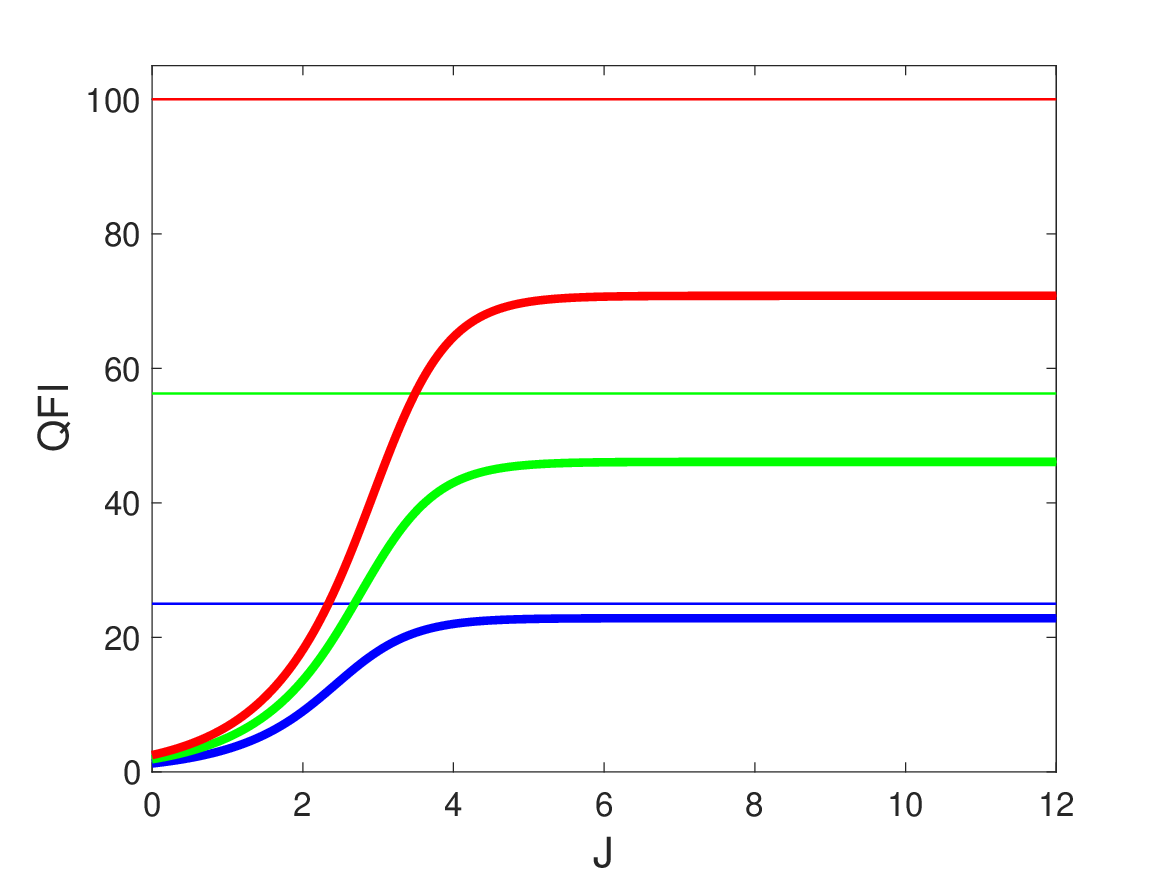}
\caption{
Dependence of the quantum Fisher information (QFI) on the coupling strength $J$. The blue, green,
and red curves correspond to system sizes $N=10$, $15$, and $20$, respectively. Thick lines represent the 
exact QFI, while thin lines indicate the corresponding bounds.
}
\label{fig:varyJ}
\end{figure}

In  Fig.~\ref{fig:varyJ}, we illustrate the dependence of the quantum Fisher information on the coupling strength $J$. 
The parameters are chosen as $B_1 = 0$, $B_2 = 0.06$, and $\beta = 0.5$.
The horizontal axis represents $J$, while the vertical axis shows $\mathbf{n}^T \mathbf{F} \mathbf{n}$ with $\mathbf{n} = (1/2, 1/2)$.
The thick curves correspond to the exact values of the quantum Fisher information, whereas the thin curves denote the corresponding upper bounds.
From bottom to top, the thick (thin) curves correspond to systems sizes  
$N=10$, $15$, and $20$, respectively. 
As expected, the bound consistently lies above the exact quantum Fisher information, in agreement with our theoretical analysis.
 
\begin{figure}[htbp]
\centering
\includegraphics[width=8.5cm]{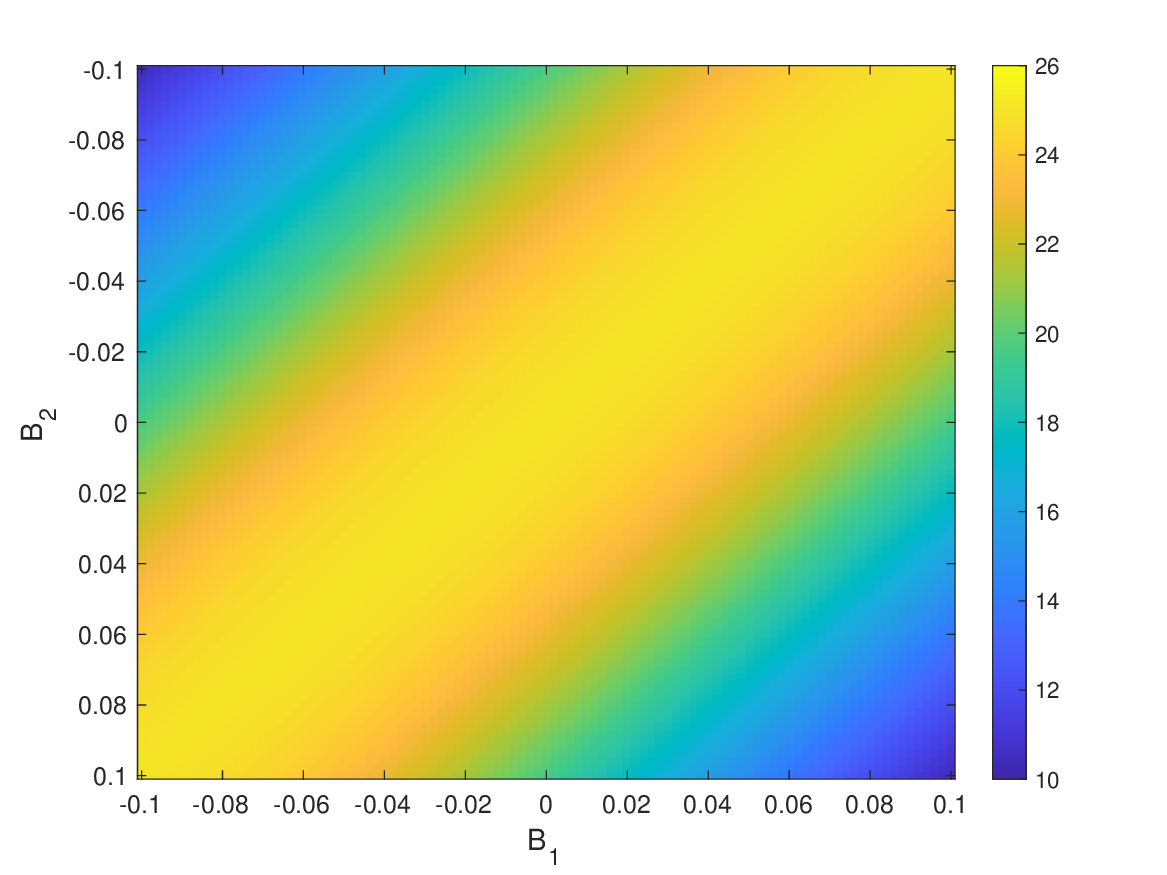}
\caption{
Dependence of the quantum Fisher information (QFI) on the local fields $B_1$ and $B_2$.
The values corresponding to the color represents the QFI values. The plot shows
that the QFI value is constant for a fixed $B_1+B_2$ value.
}
\label{fig:varyB}
\end{figure}

In Fig.~\ref{fig:varyB}, we present the dependence of the quantum Fisher information 
on the parameters $B_1$ and $B_2$. The remaining parameters are fixed
at $\beta = 0.5$, $J=6$, and $N=10$. The quantum Fisher information remains a constant for a fixed value of $B_1+B_2$. 
Moreover, the quantum Fisher information attains its maximum along the line
$B_1 + B_2 = 0$ and decreases monotonically as $|B_1+ B_2 |$ increases.

This behavior admits a simple physical interpretation. The condition $B_1 + B_2 = 0$ corresponds to a purely staggered field with vanishing uniform component. Along this line, the Hamiltonian preserves a sublattice symmetry that enhances the competition between the interaction term and the alternating field, leading to increased susceptibility of the thermal state to parameter variations. In contrast, a nonzero uniform component $B_1 + B_2$ effectively biases the system toward a globally polarized configuration, which suppresses fluctuations and reduces the QFI. Therefore, parameter combinations that minimize the uniform field component and maximize sublattice contrast are most favorable for achieving enhanced metrological sensitivity in this model.

%\section{Attainability}
\section{Attainability}

For general Hamiltonians, the associated parameters are usually incompatible, making the sensitivity bounds of QCRB often unattainable.
Consequently, it is important to understand when the bounds are unattainable. This issue is the focus of this section.
Note that the Holevo Cram\'er-Rao bound can always be attained despite possible incompatible parameters, and is at most a factor of $\sqrt{2}$ away from QCRB \cite{albarelli2019evaluting}.

A sufficient condition for attainability of QCRB \cite{crowley2014tradeoff} is $\textrm{Tr}(\rho [L_\mu, L_\nu])=0$, where 
$L_\mu$ and $L_\nu$ denote the symmetric logarithmic derivatives (SLDs). 
To understand this condition, we first express the SLDs as
$L_\mu = J_B^{-1} (\dot{\rho_\mu})$,
and
$L_\nu = J_B^{-1} (\dot{\rho_\nu})$.
The attainability condition then becomes
\begin{equation}
\textrm{Tr}(\rho [J_B^{-1} (\dot{\rho_\mu}) , J_B^{-1} (\dot{\rho_\nu})] )=0.
\end{equation}

In equilibrium quantum metrology, the probe state takes the thermal form
\begin{equation}
\rho = \frac{e^{-\beta H}}{\operatorname{Tr}\!\left[e^{-\beta H}\right]} .
\end{equation}
Its parameter derivatives are given by
\begin{equation}
\begin{aligned}
\dot{\rho}_\mu
&= - J_L\!\left[\beta \dot{H}_\mu\right]
   + \rho \operatorname{Tr}\!\left[\beta \dot{H}_\mu \rho\right], \\
\dot{\rho}_\nu
&= - J_L\!\left[\beta \dot{H}_\nu\right]
   + \rho \operatorname{Tr}\!\left[\beta \dot{H}_\nu \rho\right].
\end{aligned}
\end{equation}
Substituting these expressions into the definitions of the SLDs yields
\begin{equation}
\begin{aligned}
L_\mu
&= - J_B^{-1} J_L\!\left[\beta \dot{H}_\mu\right]
   + \operatorname{Tr}\!\left[\beta \dot{H}_\mu \rho\right] \mathbb{I}, \\
L_\nu
&= - J_B^{-1} J_L\!\left[\beta \dot{H}_\nu\right]
   + \operatorname{Tr}\!\left[\beta \dot{H}_\nu \rho\right] \mathbb{I}.
\end{aligned}
\end{equation}
Combining the above expressions, we find
\begin{equation}
\begin{aligned}
\beta^{-2} \left( L_\mu L_\nu - L_\nu L_\mu \right)
&= J_B^{-1} J_L\!\left[\dot{H}_\mu\right]
   J_B^{-1} J_L\!\left[\dot{H}_\nu\right] \\
&\quad
   - J_B^{-1} J_L\!\left[\dot{H}_\nu\right]
     J_B^{-1} J_L\!\left[\dot{H}_\mu\right].
\end{aligned}
\end{equation}  
Therefore, the attainability condition can be rewritten as
\begin{equation}
\begin{aligned}
\beta^{-2} \textrm{Tr}(\rho [L_\mu, L_\nu]) =  & \textrm{Tr}( \rho J_B^{-1}J_L[ \dot{H_\mu} ]J_B^{-1}J_L[ \dot{H_\nu} ] ) \\
   &  -\textrm{Tr}(\rho J_B^{-1}J_L[ \dot{H_\nu} ]J_B^{-1}J_L[ \dot{H_\mu} ] ).
 \end{aligned}
\end{equation}

A stronger sufficient condition for attainability is the commutativity of the SLDs, $[ L_\mu, L_\nu ] = 0$.
This condition is satisfied if the following three conditions hold simultaneously: 
 \begin{enumerate}[label=(\roman*)]
        \item $[\rho,   \dot{H_\nu} ] = 0$; 
        \item $[ \rho,   \dot{H_\mu} ] = 0$; 
        \item $[  \dot{H_\mu},   \dot{H_\nu} ] = 0$.
    \end{enumerate}
To see this, note that when $[\rho, A ] = 0$, the superoperators satisfy
$J_B^{-1}[A] = A / \rho$,
and 
$J_L[A] =  \rho A$.
Under conditions (i) and (ii), the SLDs reduce to
$L_\nu = -\beta \dot{H_\nu} +  \textrm{Tr}[\beta \dot{H_\nu} \rho] \mathbb{I}$.
and 
$L_\mu = -\beta \dot{H_\mu}  +  \textrm{Tr}[\beta \dot{H_\mu} \rho] \mathbb{I}$.
Therefore, the condition
$[ L_\mu, L_\nu ] = 0$ is equivalent to $[  \dot{H_\mu},   \dot{H_\nu} ] = 0$.

For thermal state $\rho$, the three conditions can be satisfied if the following three conditions hold:
 \begin{enumerate}[label=(\roman*)]
        \item $[H,   \dot{H_\nu} ] = 0$; 
        \item $[ H,   \dot{H_\mu} ] = 0$; 
        \item $[  \dot{H_\mu},   \dot{H_\nu} ] = 0$.
    \end{enumerate}
Indeed,
$[\rho,   \dot{H_\nu} ] = 0 \leftrightarrow [ e^{-\beta H}/\textrm{Tr}[ e^{-\beta H}],   \dot{H_\nu}]=0  
\leftrightarrow [  e^{-\beta H},   \dot{H_\nu}]=0  \leftarrow [H,   \dot{H_\nu} ] = 0$,
and an analogous argument applies to $\dot{H_\mu}$.

Next, we examine two model Hamiltonians to illustrate the attainability conditions, and also describe 
the optimal measurements necessary to attain the bounds.
The first Hamiltonian is given by
\begin{equation}
H  =  \sum_{m=1}^M  (  \sum_{k=1}^{N_m}  h_m^{(k)} )  \theta_m = \sum_{m=1}^M  H_m  \theta_m.
\end{equation}
Here, it is important that $H_i$ and $H_j$ act on different subspaces for $i\not = j$.
For this Hamiltonian, the generators commute,
$\dot{H_\mu} \dot{H_\nu} =\dot{H_\nu} \dot{H_\mu}$,
since $\dot{H_\nu}= H_i$ and $\dot{H_\mu} = H_j$ act on different subspaces.
Moreover, 
$[ H,   \dot{H_\mu} ] = 0$.
To see this, note that 
$H$ can be decomposed as $\theta \dot{H_\mu}  + B + C + \cdots$, where $\theta$ is the parameter.
$\dot{H_\mu}$ and $B$ commute because they are linear operators acting on orthogonal subspaces, and similarly for $\dot{H_\mu}$ and $C$, etc.
Moreover, $[\theta A, A] = \theta A^2 -  \theta A^2 = 0$. Therefore, all relevant commutation relations vanish, and the multiparameter precision bounds can be simultaneously attained for this Hamiltonian.

The second Hamiltonian takes the form
\begin{equation}
 H =  \theta_1 \sigma_x +  \theta_2 \sigma_y +  \theta_3 \sigma_z.
 \end{equation}
In this case, 
$[ L_\mu, L_\nu ] =0$ is not satisfied since $[\sigma_x,\sigma_z ] \not =0$. 
Nevertheless, the weaker condition $\textrm{Tr}(\rho [L_\mu, L_\nu])=0$ still holds, implying that
the precision limits for all parameters remain attainable.
The reason is as follows.
First of all, the commutation relations
$[\rho,   \dot{H_\nu} ] = 0$
and
$[ \rho,   \dot{H_\mu} ] = 0$
are satisfied for this Hamiltonian.
Under these conditions,
$\textrm{Tr}(\rho [L_\mu, L_\nu])=0$ is equivalent to 
\begin{equation}
\label{eq:rHH}
\textrm{Tr}(\rho [ \dot{H_\mu},   \dot{H_\nu}])=0.
\end{equation}
Finally, the relation Eq.~\eqref{eq:rHH} holds because
$\textrm{Tr}(\rho \dot{H_\mu}  \dot{H_\nu} ) = \textrm{Tr}(  \dot{H_\mu}  \rho \dot{H_\nu} ) = \textrm{Tr}(   \rho \dot{H_\nu}  \dot{H_\mu} )$,
for all $\mu$ and $\nu$.
Hence, despite the noncommutativity of the generators, the multiparameter bounds can still be simultaneously achieved for this Hamiltonian.

Finally, we examine the optimal measurements required to reach the sensitivity bound when the attainability condition holds.
If $[ L_\mu, L_\nu ] =0$ for all $\mu$ and $\nu$, the operators $L_\mu$ share the same eigenstates for all $\mu$.
In this case, the optimal measurements are projective measurements onto the eigenstates of $L_\mu$, which are given by
\begin{equation}
\begin{aligned}
 J_B^{-1}[\dot{\rho_\mu}] |_{\mu=0} &= \sum_{ij} \frac{2(p_i-p_j)}{(\ln p_i - \ln p_j)(p_i + p_j)} H'_{ij} \ket{i} \bra{j}  \\
  & = \sum_{ij} \frac{2\tanh(\beta (E_i - E_j)/2)}{\beta (E_i-E_j)} H'_{ij} \ket{i} \bra{j},
\end{aligned}
\end{equation}
where $\rho = \sum p_i \ket{i}\bra{i}$, and we use $H' \equiv \dot{H_\mu}$ for notational simplicity.
For the case of weak commutativity, i.e., $\textrm{Tr}(\rho [L_\mu, L_\nu])=0$, we leave the determination of optimal measurements as future work.

\section{Conclusion}

In summary, we have shown that multi-parameter quantum sensing with thermal states can achieve the Heisenberg limit with respect to the number of probes
 for closed systems.
We considered two cases: the finite temperature case and the zero temperature case. The quantum Fisher information scales quadratically with the inverse temperature in the former case, and quadratically with the inverse of the energy gap in the latter case. The quantum Fisher information in both cases
scales quadratically with the resources, i.e., the number of probes. We further provide the condition 
for a sensing protocol to attain this Heisenberg limit, in particular highlighting the role of commuting generators. Several examples were provided to illustrate these 
main results.

There are several promising directions for future research. First, it would be interesting to incorporate the effects of noise into 
multi-parameter equilibrium metrology, as noise is inevitable in realistic sensing scenarios.
Note however that, compared to dynamic metrology, noise is much less significant an issue in equilibrium metrology as the probe state is in thermal equilibrium.
Hence, the QFI bound is inherently robust against noise.
Second, in this work, we have focused on closed systems.
It would also be interesting to consider multiparameter equilibrium metrology for open systems, i.e., the parameters to be estimated 
appears in a Lindbladian. 
Third, privacy considerations in multi-parameter quantum sensing with thermal states merit further investigation.
Since the estimated parameters are sometimes the private properties of their owners, it is desirable for the sensing protocol to provide
privacy guarantees \cite{hassani2025privacy}. Finally, an experimental demonstration that attains the Heisenberg limit provided in our bound would 
be an important next step toward the practical deployment of thermal states for multi-parameter quantum sensing.

% Future work

\begin{acknowledgements}
This work was supported by the Natural Science Foundation of Shanghai (25ZR1402098). 
\end{acknowledgements}

\appendix 

\section{The inverse operator $J_{B,\rho}^{-1}$}
\label{app:A}

To verify that $J_{B,\rho}^{-1}$ is indeed the inverse operator of $J_{B,\rho}$, we first give the definition of the inverse operator.
If for any bounded, invertible map $A$ on $\mathcal{B}(\mathcal{H})$, it holds that
\begin{equation}
J_{B,\rho}[J_{B,\rho}^{-1}[A]]= A,
\end{equation}
then $J_{B,\rho}^{-1}$ is called the inverse operator of $J_{B,\rho}$.

Next note that
\begin{equation}
\begin{aligned}
  &\rho  \int_0^\infty ds e^{-\rho s} A e^{-\rho s} +  \int_0^\infty ds e^{-\rho s} A e^{-\rho s} \rho  \\
=  & \int_0^\infty ds \rho e^{-\rho s} A e^{-\rho s} +  \int_0^\infty ds e^{-\rho s} A e^{-\rho s} \rho \\
= & -  \int_0^\infty ds ( \frac{d}{ds} e^{-s\rho} A e^{ -s \rho} ) = ( e^{-s \rho} A e^{ -s \rho} )|^{s=0}_{s=\infty} = A,
\end{aligned}
\end{equation}
where the integral converges absolutely in operator norm due to the strict positivity of $\rho$.
This confirms that $J_{B,\rho}^{-1}$ acts as the inverse of $J_{B,\rho}$.

\section{Proof of Eq.~\eqref{eq:thermal}}
\label{app:B}

To prove Eq.~\eqref{eq:thermal}, we begin with Eq.~\eqref{eq:QFI}.  
Using the result derived in Ref.~\cite{abiuso2025fundamental}, the derivatives $ \dot{\rho_\mu}$ and  $ \dot{\rho_\nu}$ can be written as
\begin{equation}
\label{eq:2}
 \dot{\rho_\mu} = - J_{L,\rho} [\beta \dot{H_\mu}] + \rho {\rm Tr} [ \beta \dot{H_\mu} \rho ],
\end{equation}
and 
\begin{equation}
\label{eq:3}
 \dot{\rho_\nu} = - J_{L,\rho} [\beta \dot{H_\nu}] + \rho {\rm Tr} [ \beta \dot{H_\nu} \rho ].
\end{equation}

Substituting Eqs.~\eqref{eq:2} and \eqref{eq:3} into Eq.~\eqref{eq:QFI}, we obtain
\begin{equation}
\begin{aligned}
 \beta^{-2} F_{\mu, \nu} = &  {\rm Tr} [  J_{L, \rho} [ \dot{H_\mu} ]  J_{B,\rho}^{-1}  \circ J_{L, \rho} [ \dot{H_\nu} ]  ]  \\
 &  + {\rm Tr} [ \rho  J_{B,\rho}^{-1} [\rho]  ]     {\rm Tr}[  \rho \dot{H_\mu}  ]       {\rm Tr}[  \rho \dot{H_\nu}  ]          \\
 & -  {\rm Tr}  [  J_{L, \rho} [ \dot{H_\mu} ]   J_{B,\rho}^{-1} [\rho] ]       {\rm Tr}[  \rho \dot{H_\nu}  ]    \\
  &  - {\rm Tr} [ J_{B,\rho}^{-1} [\rho]    J_{L, \rho} [ \dot{H_\nu} ]     ]         {\rm Tr}[  \rho \dot{H_\mu}  ] .
\end{aligned}
\end{equation}
Since $ J_{B,\rho}^{-1} [\rho] = \mathbbm{1}$, this expression simplifies to 
\begin{equation}
\begin{aligned}
 \beta^{-2} F_{\mu, \nu} = &  {\rm Tr} [  J_{L, \rho} [ \dot{H_\mu} ]  J_{B,\rho}^{-1}  \circ J_{L, \rho} [ \dot{H_\nu} ]  ]   +    {\rm Tr}[  \rho \dot{H_\mu}  ]       {\rm Tr}[  \rho \dot{H_\nu}  ]                                   \\
 & -  {\rm Tr}  [  J_{L, \rho} [ \dot{H_\mu} ]   ]       {\rm Tr}[  \rho \dot{H_\nu}  ]        - {\rm Tr} [ J_{L, \rho} [ \dot{H_\nu} ]     ]         {\rm Tr}[  \rho \dot{H_\mu}  ]  .
\end{aligned}
\end{equation}

Finally, using the identities  ${\rm Tr} [ J_{L, \rho} [ A ]  ]  = {\rm Tr} [  \rho A  ] $
and $ {\rm Tr} [ J_{L, \rho} [ A ]  B ] = {\rm Tr} [ A J_{L, \rho} [ B ]  ]$, we have
\begin{equation}
 \beta^{-2} F_{\mu, \nu} =   {\rm Tr} [   \dot{H_\mu}  \mathcal{J}_\rho [ \dot{H_\nu} ]  ]    -   {\rm Tr}[  \rho \dot{H_\mu}  ]       {\rm Tr}[  \rho \dot{H_\nu}  ]    ,   
\end{equation}
where $ \mathcal{J}_\rho = J_{L,\rho} \circ J_{B,\rho}^{-1} \circ J_{L,\rho}$.

\section{Proof of Eq.~\eqref{eq:thermal_bound}}
\label{app:C}

Let $ G =  \beta^{-2} F$. It suffices to prove that
\begin{equation}
\mathbf{n}^T G \mathbf{n} \le \mathbf{n}^T \Gamma \mathbf{n}.
\end{equation}
Expanding both sides yields
\begin{equation}
\sum\limits_{i,j=1}^M G_{ij} n_i n_j  \le \sum\limits_{i,j=1}^M \Gamma_{ij} n_i n_j.
\end{equation}
Substituting the expression of $G_{ij}$ and $\Gamma_{ij}$ and eliminating the common term, we have
\begin{equation}
\label{eq:mid}
\begin{aligned}
\sum\limits_{i,j=1}^M  n_i n_j  {\rm Tr} [ \dot{H_i} \mathcal{J}_\rho [ \dot{H_j}] ]  &  \le \sum\limits_{i,j=1}^M  n_i n_j    {\rm Tr}[  \rho \frac{ \dot{H_\mu} \dot{H_\nu} + \dot{H_\nu} \dot{H_\mu} }{2}   ] \\
& = \sum\limits_{i,j=1}^M  n_i n_j    {\rm Tr}[ \dot{H_i} J_{B,\rho}[\dot{H_j}]  ],
\end{aligned}
\end{equation}
where the Bogoliubov operator is defined as $J_{B,\rho}[A] = (\rho A + A \rho)/2$.

To establish this inequality, we introduce some additional notations. Let $A^i_{sr}$ denote the $(s,r)$-th entry of the Hermitian matrix $\dot{H_i}$, and similarly for $A^j_{sr}$.
Write the spectral decomposition of  $\rho$ as
\begin{equation}
\rho = \sum_i p_i \ket{i}\bra{i},
\end{equation}
where $\ket{i}$ are the eigenstates of $\rho$.
Define the coefficients
\begin{equation}
\begin{aligned} 
a_{sr} &  = \frac{2(p_s-p_r)^2 }{ ( \ln p_s -\ln p_r  )^2 (p_s + p_r)},  \\
b_{sr} & =  \frac {p_s + p_r } {2}.
\end{aligned}
\end{equation}
According to Ref.~\cite{abiuso2025fundamental},  the actions of $J_{B,\rho}$ and
$\mathcal{J}_\rho$ on the operator basis $\{ \ket{i}\bra{j} \}$ are diagonal,
\begin{equation}
\begin{aligned} 
J_{B,\rho}[ \ket{i}\bra{j} ] & = b_{ij} \ket{i}\bra{j},  \\
\mathcal{J}_\rho [ \ket{i}\bra{j} ] & =  a_{ij} \ket{i}\bra{j}.
\end{aligned}
\end{equation}

Substituting these expressions into Eq.~\eqref{eq:mid} gives
\begin{equation}
\sum\limits_{i,j=1}^M n_i n_j  \sum\limits_{s,r=1}^D (A^i_{sr})^*  A^j_{sr} a_{sr} \le  \sum\limits_{i,j=1}^M n_i n_j  \sum\limits_{s,r=1}^D (A^i_{sr})^*  A^j_{sr} b_{sr}.
\end{equation}
Reordering the summation, we get
\begin{equation}
 \sum\limits_{s,r=1}^D  a_{sr}  \sum\limits_{i,j=1}^M n_i n_j   (A^i_{sr})^*  A^j_{sr} \le   \sum\limits_{s,r=1}^D  b_{sr}  \sum\limits_{i,j=1}^M n_i n_j   (A^i_{sr})^*  A^j_{sr}.
 \end{equation}
 Noting that
 \begin{equation}
  | \sum\limits_{i=1}^M n_i A^i_{sr} |^2 \ge 0,
\end{equation}
 the inequality reduces to
\begin{equation}
 \sum\limits_{s,r=1}^D  a_{sr}  | \sum\limits_{i=1}^M n_i A^i_{sr} | ^ 2  \le    \sum\limits_{s,r=1}^D  b_{sr}  | \sum\limits_{i=1}^M n_i A^i_{sr} | ^ 2.
\end{equation} 
This inequality holds because $a_{sr} \le b_{sr}$ for all $(s,r)$, as proven in Ref.~\cite{abiuso2025fundamental}, and each summand is non-negative.
This finishes the proof.

\section{Proof of Eq.~\eqref{eq:GammaHL} and Eq.~\eqref{eq:GammaSat}}
\label{app:GammaOpt}
Since  $\Gamma(\rho,H)$ is the covariance matrix of the Hamiltonian generators,
\begin{equation}
\Gamma_{\mu\nu}(\rho,H)
=
\frac{1}{2}\langle H_\mu H_\nu + H_\nu H_\mu\rangle_\rho
-
\langle H_\mu\rangle_\rho \langle H_\nu\rangle_\rho,
\end{equation}
we have
\begin{equation}
\mathbf{n}^T \Gamma(\rho,H)\mathbf{n}
=
\mathrm{Var}_\rho(H_{\mathbf{n}}),
\qquad
H_{\mathbf{n}} := \sum_{k=1}^M n_k H_k .
\end{equation}
The maximization over all density matrices $\rho$ therefore reduces to maximizing the variance of the collective observable $H_{\mathbf{n}}$.  

For any Hermitian operator $A$, the maximal variance over all quantum states is achieved by a pure state supported on the extremal eigenspaces of $A$. If $\lambda_{\max}$ and $\lambda_{\min}$ denote the largest and smallest eigenvalues of $A$, then
\begin{equation}
\max_\rho \mathrm{Var}_\rho(A)
=
\frac{(\lambda_{\max}-\lambda_{\min})^2}{4},
\end{equation}
and the maximum is reached by the equal superposition
\begin{equation}
\frac{1}{\sqrt{2}}
\left(
\ket{\lambda_{\max}}
+
\ket{\lambda_{\min}}
\right).
\end{equation}
This follows from the convexity of the variance in $\rho$ (so the maximum is attained for pure states) and from the fact that the variance of a pure state is bounded by the spectral diameter of $A$.
Applying this result to $A = H_{\mathbf{n}}$, the problem reduces to determining the spectral range of $H_{\mathbf{n}}$.

\subsection{Spectral range of $H_{\mathbf{n}}$.}

Each generator has the structure
\begin{equation}
H_k
=
\epsilon_k \sum_{i=1}^{N_k} h_{k,i},
\qquad
\epsilon_k = \mathrm{sgn}(n_k),
\end{equation}
where $h_{k,i}$ has eigenvalues $\pm 1/2$.  
Hence the maximal and minimal eigenvalues of $H_k$ are
$\pm N_k/2$.
It follows that
\begin{equation}
H_{\mathbf{n}}
=
\sum_{k=1}^M |n_k|
\sum_{i=1}^{N_k} h_{k,i},
\end{equation}
whose extremal eigenvalues are obtained when all spins are aligned:
\begin{equation}
\lambda_{\max}
=
\frac{1}{2}\sum_{k=1}^M |n_k| N_k,
\qquad
\lambda_{\min}
=
-
\frac{1}{2}\sum_{k=1}^M |n_k| N_k.
\end{equation}
Therefore,
\begin{equation}
\max_\rho
\mathbf{n}^T \Gamma(\rho,H)\mathbf{n}
=
\frac{1}{4}
\left(
\sum_{k=1}^M |n_k| N_k
\right)^2.
\end{equation}

Defining the vector
\begin{equation}
\boldsymbol{v}
=
(\epsilon_1 N_1,\ldots,\epsilon_M N_M),
\end{equation}
we obtain
\begin{equation}
\Gamma_{HL}
=
\frac{\boldsymbol{v}\boldsymbol{v}^T}{4},
\end{equation}
so that
\begin{equation}
\mathbf{n}^T \Gamma(\rho,H)\mathbf{n}
\le
\mathbf{n}^T \Gamma_{HL} \mathbf{n}.
\end{equation}

\subsection{Saturation by the GHZ-like state.}

The upper bound is saturated by a pure state that is an equal superposition of the eigenstates corresponding to $\lambda_{\max}$ and $\lambda_{\min}$.  

The states
\begin{equation}
\ket{N_m,1},
\qquad
\ket{N_m,-1},
\end{equation}
are fully polarized states of the $m$-th block, with all local eigenvalues equal to $+1/2$ or $-1/2$, respectively. The tensor product
\begin{equation}
\ket{N_1,\epsilon_1}\otimes\cdots\otimes\ket{N_M,\epsilon_M}
\end{equation}
is therefore the eigenstate of $H_{\mathbf{n}}$ with eigenvalue $\lambda_{\max}$, while
\begin{equation}
\ket{N_1,-\epsilon_1}\otimes\cdots\otimes\ket{N_M,-\epsilon_M}
\end{equation}
has eigenvalue $\lambda_{\min}$.

The GHZ-like state in Eq.~\eqref{eq:GammaSat},
\begin{equation}
\ket{\psi}
=
\frac{1}{\sqrt{2}}
\left(
\ket{\lambda_{\max}}
+
\ket{\lambda_{\min}}
\right),
\end{equation}
is precisely the equal superposition of these two extremal eigenstates. For this state,
\begin{equation}
\langle H_{\mathbf{n}} \rangle = 0,
\qquad
\langle H_{\mathbf{n}}^2 \rangle
= \lambda_{\max}^2,
\end{equation}
so that
\begin{equation}
\mathrm{Var}_{\psi}(H_{\mathbf{n}})
=
\lambda_{\max}^2
=
\frac{1}{4}
\left(
\sum_{k=1}^M |n_k| N_k
\right)^2.
\end{equation}
Thus the spectral-diameter bound is saturated, yielding
\begin{equation}
\mathbf{n}^T \mathbf{F}(\rho,H)\mathbf{n}
\le
\beta^2
\mathbf{n}^T \Gamma_{HL}\mathbf{n},
\end{equation}
with equality achieved by the GHZ-like state.

\section{Thermal QFI Bounds in Continuous-Variable Systems}
 \label{app:CV}

In this section, we extend the upper bounds of the quantum Fisher information (QFI) matrix for multi-parameter estimation at thermal equilibrium to continuous-variable (CV) systems.

We consider a bosonic system composed of $K$ modes, with canonical quadrature operators collected into the vector
\begin{equation}
\mathbf{R} = (\hat{x}_1, \hat{p}_1, \dots, \hat{x}_K, \hat{p}_K)^T,
\end{equation}
satisfying the canonical commutation relations $[\hat{x}_k,\hat{p}_\ell]=i\delta_{k\ell}$ (we set $\hbar=1$).

The Hamiltonian is assumed to encode $M$ parameters locally as
\begin{equation}
H = \sum_{m=1}^M \left( \sum_{k=1}^{N_m} \hat{g}_m^{(k)} \right) \theta_m,
\end{equation}
where $\hat{g}_m^{(k)}$ are Hermitian operators acting on individual modes (or finite subsets of modes), and $\theta_m$ are the unknown parameters.

We consider the Gibbs state at inverse temperature $\beta$,
\begin{equation}
\rho = \frac{e^{-\beta H}}{Z}, \quad Z = \mathrm{Tr}(e^{-\beta H}),
\end{equation}
which is well-defined provided $H$ is bounded from below and $Z<\infty$. For quadratic Hamiltonians, $\rho$ is a Gaussian state fully characterized by its first and second moments.

In contrast to finite-dimensional systems, the operators $\hat{g}_m^{(k)}$ are generally unbounded. Therefore, instead of assuming spectral boundedness, we impose an energy constraint on the state $\rho$:
\begin{equation}
\mathrm{Tr}[\rho \hat{n}_k] \le \bar{n}, \quad \forall k,
\end{equation}
where $\hat{n}_k = a_k^\dagger a_k$ is the number operator of mode $k$, and $\bar{n}$ is a finite constant. This ensures that all relevant moments are well-defined.

To bound the QFI matrix, we introduce the covariance matrix $\Gamma(\rho,H)$ associated with the state $\rho$ and the Hamiltonian $H$. Its $(\mu,\nu)$-th entry is defined as
\begin{equation}
\Gamma(\rho,H)_{\mu,\nu}
= \frac{1}{2}\mathrm{Tr}\big[\rho(\dot{H}_\mu \dot{H}_\nu + \dot{H}_\nu \dot{H}_\mu)\big]
- \mathrm{Tr}[\rho \dot{H}_\mu]\, \mathrm{Tr}[\rho \dot{H}_\nu],
\end{equation}
where $\dot{H}_\mu = \partial H / \partial \theta_\mu = \sum_{k=1}^{N_\mu} \hat{g}_\mu^{(k)}$.

For any real unit vector $\mathbf{n}=(n_1,\dots,n_M)$ with $||\mathbf{n}||_2=1$, the following inequality holds:
\begin{equation}
\label{eq:thermal_bound_cv}
\mathbf{n}^T \mathbf{F}(\rho,H) \mathbf{n}
\le \beta^2 \mathbf{n}^T \Gamma(\rho,H) \mathbf{n},
\end{equation}
which provides an upper bound on the QFI matrix. This inequality remains valid in the CV setting provided all second moments are finite.

We now derive a more explicit bound on the quadratic form $\mathbf{n}^T \Gamma(\rho,H) \mathbf{n}$. In contrast to finite-dimensional systems, the operators $\hat{g}_m^{(k)}$ are not norm-bounded. Instead, their fluctuations are controlled by the energy constraint.

Assuming that each generator satisfies
\begin{equation}
\mathrm{Tr}[\rho\, (\hat{g}_m^{(k)})^2] \le c\, (\bar{n}+1),
\end{equation}
for some constant $c>0$, we obtain
\begin{equation}
\Gamma(\rho,H)_{\mu,\nu}
\le \delta_{\mu\nu} \, c\, N_\mu (\bar{n}+1),
\end{equation}
which implies
\begin{equation}
\label{eq:Gamma_bound_cv}
\mathbf{n}^T \Gamma(\rho,H) \mathbf{n}
\le c (\bar{n}+1) \sum_{m=1}^M n_m^2 N_m.
\end{equation}

Combining Eq.~\eqref{eq:thermal_bound_cv} and Eq.~\eqref{eq:Gamma_bound_cv}, we arrive at
\begin{equation}
\label{eq:QFI_bound_cv}
\mathbf{n}^T \mathbf{F}(\rho,H) \mathbf{n}
\le c\, \beta^2 (\bar{n}+1) \sum_{m=1}^M n_m^2 N_m.
\end{equation}

For $M=1$, Eq.~\eqref{eq:QFI_bound_cv} reduces to
\begin{equation}
F \le c\, \beta^2 (\bar{n}+1) N,
\end{equation}
which exhibits linear scaling in the number of modes $N$ under fixed energy per mode.

For the multi-parameter case with $N_1=\cdots=N_M=N$, the bound implies
\begin{equation}
\mathbf{n}^T \Sigma \mathbf{n}
\ge O\!\left(\frac{1}{\beta^2 (\bar{n}+1) N}\right),
\end{equation}
which replaces the $O(1/N^2)$ scaling of finite-dimensional systems with an energy-constrained scaling.

\section{Proof of Eq.~\eqref{eq:middle}}
\label{app:D}

We aim to prove that
\begin{equation}
\begin{aligned}
 {\rm Tr} [B \mathcal{J}_\rho [ A ]  ]  =  & \sum_a  a {\rm Tr}[B \Pi_a A \Pi_a]    \\
 & +  \sum_{a \not= b} \frac{ 2(a-b)^2 }{ ( \ln a - \ln b)^2 (a+b) }  {\rm Tr}[B \Pi_b A \Pi_a].
\end{aligned}
\end{equation}
Here $\Pi_a = \ket{a}\bra{a}$ denotes the projector onto the eigenspace of the state
$\rho$ corresponding to eigenvalue $a$, and the projectors satisfy
$ \sum_a \Pi_a = \mathbbm{1}$.

Using the spectral decomposition of $\rho$ and the linearity of $\mathcal{J}_\rho$,
we first note that
\begin{equation}
 \mathcal{J}_\rho [\Pi_a  A \Pi_a]  =  a \Pi_a A \Pi_a, 
\end{equation}
and, for $a \neq b$,
\begin{equation}
 \mathcal{J}_\rho [\Pi_a  A \Pi_b]  =  \frac{2(a-b)^2}{(\ln a - \ln b)^2(a+b) } \Pi_a  A \Pi_b, 
\end{equation}

We may therefore expand $A$ in the eigenbasis of $\rho$ as
\begin{equation}
 A = \left( \sum_a \Pi_a \right) A \left( \sum_a \Pi_a \right),
\end{equation}
which yields
\begin{equation}
\begin{aligned}
 {\rm Tr} [B \mathcal{J}_\rho [ A ]  ]  
 =& {\rm Tr} [B \mathcal{J}_\rho [ (\sum_a \Pi_a) A  (\sum_a \Pi_a)]  ]   \\
= &   \sum_a  a {\rm Tr}[B \Pi_a A \Pi_a]   \\
  & +  \sum_{a \not= b} \frac{ 2(a-b)^2 }{ ( \ln a - \ln b)^2 (a+b) }  {\rm Tr}[B \Pi_b A \Pi_a],
 \end{aligned}
\end{equation}
which establishes Eq.~\eqref{eq:middle}.

\section{Proof of Eq.~\eqref{eq:Fexp}}
\label{app:E}

The derivation is as follows,
\begin{equation}
\begin{aligned}
 \frac{\partial^2}{\partial \mu \partial \nu} \ln \mathcal{Z} = &\frac{\mathcal{Z}_{\mu \nu}}{\mathcal{Z}} -  \frac{\mathcal{Z}_{\mu}}{\mathcal{Z}} \cdot  \frac{\mathcal{Z}_{\nu}}{\mathcal{Z}} \\
 = &\frac{ -\beta \frac{\partial}{\partial \mu} [exp(-\beta H)] H_\nu }{\mathcal{Z}} -  (-\beta \frac{Tr[e^{-\beta H} H_\mu] }{\mathcal{Z}})  \\
  & \cdot (-\beta \frac{Tr[e^{-\beta H} H_\nu] }{\mathcal{Z}}) \\
 =  &\beta^2 \left(Tr[\rho  H_\mu H_\nu] -  Tr[\rho H_\mu]  Tr[\rho H_\nu] \right) = F_{\mu \nu}.
 \end{aligned}
 \end{equation}

%%%%%%%%%%%%%%%%%%%%%%%%%%%%%%%%%%%%%%%%
% choose a style
%\bibliographystyle{ieeetr}
%\bibliographystyle{unsrt}
%\bibliographystyle{apsrev4-2}
% \bibliographystyle{iopart-num}
%%%%%%%%%%%%%%%%%%%%%%%%%%%%%%%%%%%%%%%%

%%%%%%%%%%%%%%%%%%%%%%%%%%%%%%%%%%%%%%%%
% choose a .bib file
\bibliography{BibliMetrology}
%%%%%%%%%%%%%%%%%%%%%%%%%%%%%%%%%%%%%%%%

\end{document}